\documentclass[sigconf,screen]{acmart}
\usepackage{enumitem}
\AtBeginDocument{%
  }

\setcopyright{none} 
\renewcommand\footnotetextcopyrightpermission[1]{}
\usepackage[normalem]{ulem}

\begin{document}
\title{Knowledge-Based Multi-Agent Framework for Automated Software Architecture Design}

\author{Yiran Zhang}
\email{yiran002@e.ntu.edu.sg}
\affiliation{
  \institution{Nanyang Technological University}
  % \city{Singapore}
  \country{Singapore}
}

\author{Ruiyin Li}
\email{ryli_cs@whu.edu.cn}
\affiliation{%
  \institution{Wuhan University}
  \city{Wuhan}
  \country{China}
}

\author{Peng Liang}
\email{liangp@whu.edu.cn}
\affiliation{
  \institution{Wuhan University}
  \city{Wuhan}
  \country{China}
}

\author{Weisong Sun}
\authornote{Weisong Sun is the corresponding author.}
\email{weisong.sun@ntu.edu.sg}
\affiliation{
  \institution{Nanyang Technological University}
  % \city{Singapore}
  \country{Singapore}
}

\author{Yang Liu}
\email{yangliu@ntu.edu.sg}
\affiliation{
  \institution{Nanyang Technological University}
  % \city{Singapore}
  \country{Singapore}
}

\begin{abstract}
Architecture design is a critical step in software development. However, creating a high-quality architecture is often costly due to the significant need for human expertise and manual effort. Recently, agents built upon Large Language Models (LLMs) have achieved remarkable success in various software engineering tasks. Despite this progress, the use of agents to automate the architecture design process remains largely unexplored.  
To address this gap, we envision a Knowledge-based Multi-Agent Architecture Design (MAAD) framework. MAAD uses agents to simulate human roles in the traditional software architecture design process, thereby automating the design process. To empower these agents, MAAD incorporates knowledge extracted from three key sources: 1) existing system designs, 2) authoritative literature, and 3) architecture experts. By envisioning the MAAD framework, we aim to advance the full automation of application-level system development.
\end{abstract}

%   <concept_desc>Software and its engineering~Software development techniques</concept_desc>

% \ccsdesc[500]{Software and its engineering~Designing software}
% \ccsdesc[500]{Software and its engineering~Collaboration in software development}
% \ccsdesc[500]{Human-centered computing~Interaction design}


\received{20 February 20xx}
\received[revised]{12 March 20xx}
\received[accepted]{5 June 20xx}

\maketitle

\section{Introduction}\label{sec:intro}
Software architecture plays a vital role in the software development process. It serves as the blueprint that ensures systems are scalable, maintainable, and aligned with the business goals specified in the Software Requirements Specification (SRS)~\cite{Bass2012SAP}. Traditionally, the architecture design process relies heavily on human expertise. This manual approach increases costs and makes the process time-consuming and inconsistent~\cite{Garlan2009ArchMism}. Therefore, automating architectural design is essential for achieving effective and efficient automation of end-to-end application-level software development.

% Large language models (LLMs) have been widely used in various software engineering areas to 
Many attempts have been made to automate architecture design. Recent studies on code generation have begun to incorporate the architectural design process~\cite{hong2023metagpt, islam2024mapcoder, lin2024llm}. However, these efforts primarily rely on a straightforward zero-shot prompting strategy for architecture design. As highlighted by Dhar \textit{et al}.~\cite{dhar2024can}, such approaches frequently lead large language models (LLMs) to generate hallucinatory content. A promising solution to this issue is the adoption of multi-agent systems, which leverage interactions among multiple agents to enhance reliability and creativity, effectively mitigate hallucinations, and improve overall output quality~\cite{du2023improving, talebirad2023multi}. However, whether a multi-agent system can be used to automate the architecture design process remains unexplored. 
% Many attempts have been made to automate architecture design. Recent studies on code generation have begun to incorporate the architectural design process~\cite{hong2023metagpt, islam2024mapcoder, lin2024llm}. However, these efforts primarily rely on a straightforward zero-shot prompting strategy for architecture design. As highlighted by Dhar \textit{et al}.~\cite{dhar2024can}, such approaches frequently lead large language models (LLMs) to generate hallucinatory content. A promising solution to this issue is the adoption of multi-agent systems, which leverage interactions among multiple agents to enhance reliability and creativity, effectively mitigate hallucinations, and improve overall output quality~\cite{du2023improving, talebirad2023multi}. However, whether a multi-agent system can be used to automate the architecture design process remains unexplored. 

\begin{figure*}[hbtp]
	\centering
        \includegraphics[width=0.95\textwidth]{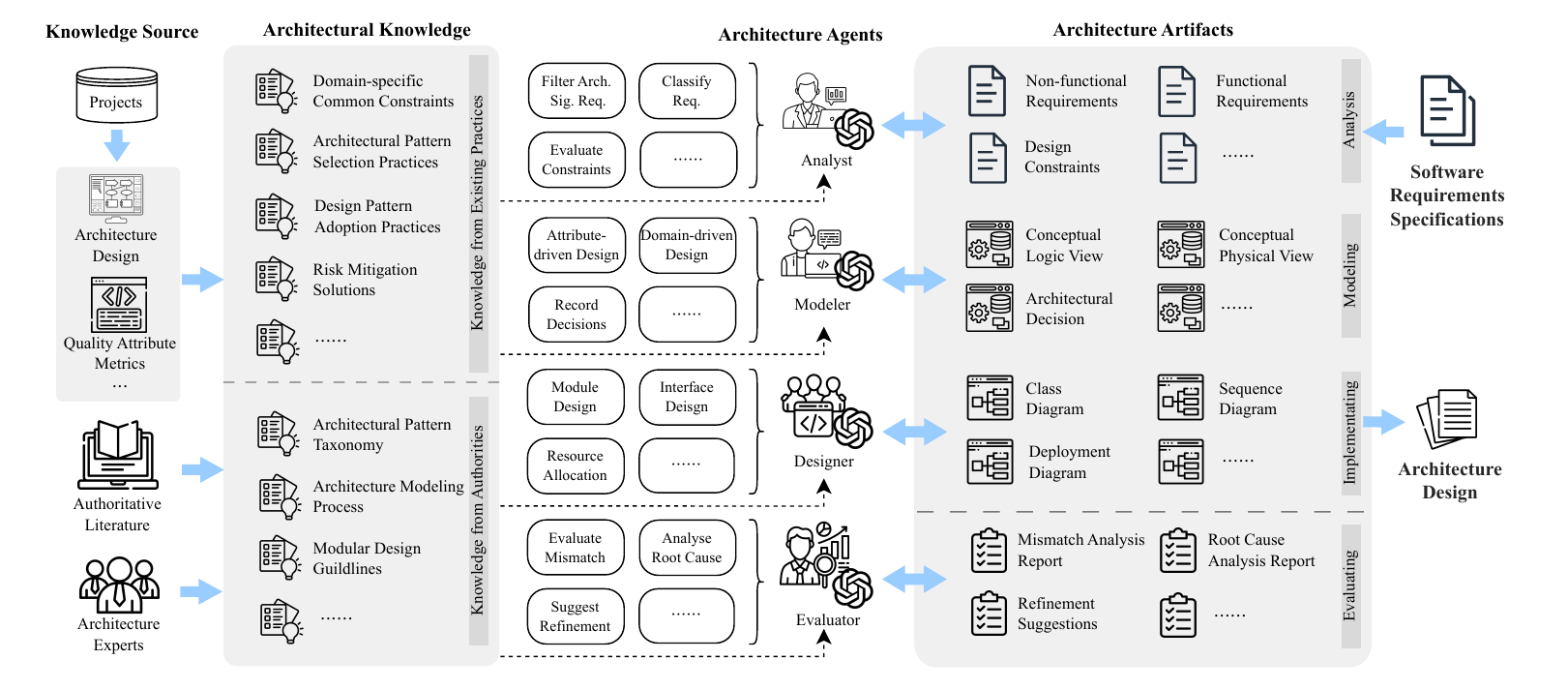}
	\caption{The Overview of Knowledge-Based Multi-Agent Framework for Architecture Design}\label{F:Overview}
\end{figure*}

To fill this gap, in this paper, we present our vision for an automated software architecture design framework, \textit{Multi-Agent Architecture Design (MAAD)}. The overview of MAAD is shown in Figure~\ref{F:Overview}. MAAD involves collaboration among four architecture agents to jointly implement the system architecture design based on the input SRS. Additionally, to support these agents, we propose potential solutions for extracting architectural knowledge from both existing system designs and authoritative sources. By addressing key challenges and opportunities in automated architecture design, we aim to advance the automation of end-to-end application-level software development. 

\section{MAAD Framework}\label{Sec:MAAD}

In this section, we will detail our vision of MAAD as shown in Figure~\ref{F:Overview}. The four agents, Analyst, Modeler, Designer, and Evaluator, will collaboratively design the system architecture based on the input SRS. First, we will illustrate the collaboration process. Then, we will provide a detailed description of each agent's design.

\subsection{Collaboration Process}
\begin{figure*}[hbtp]
	\centering
        \includegraphics[width=0.90\textwidth]{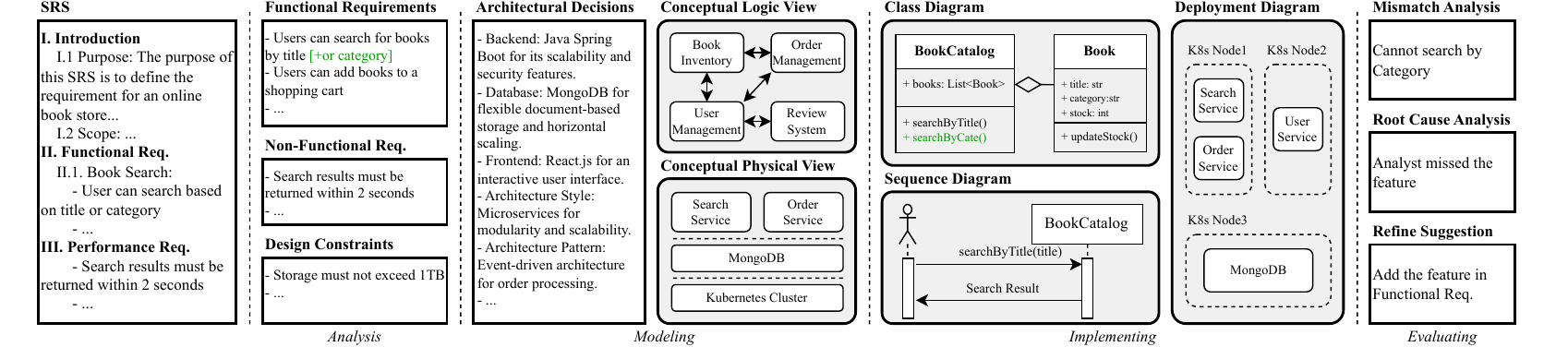}
	\caption{An Example of Proposed Architecture Design Process}\label{F:Case}
\end{figure*}

To facilitate understanding, an example of developing an online bookstore is shown in Figure~\ref{F:Case}. 
Initially, the Analyst agent analyzes the input SRS by extracting requirements and constraints that influence the architecture design. 
Building on the Analyst agent's analysis results, the Modeler agent undertakes two primary tasks: 1) formulating high-level architectural decisions to guide the design, and 2) specifying the domains and prioritized quality attributes that the generated system needs to address. Next, the Designer agent conducts concrete architecture implementation by creating UML diagrams (class diagrams, sequence diagrams, and deployment diagrams, etc.), which serve as blueprints to guide follow-up code development. 
Finally, the Evaluator agent verifies whether the artifacts align with the input SRS. If any mismatches are found, it identifies the root cause and collaborates with the relevant agent to resolve the issue. The remaining agents then update their respective artifacts accordingly. 
The architecture design process concludes when the Evaluator confirms the artifacts. The final architecture design, including the designed diagrams, conceptual views, and documented architectural decisions, collectively serves as the foundation for the subsequent code implementation.

\subsection{Architecture Agents}
\subsubsection{Analyst}

The Analyst agent is responsible for understanding the SRS, filtering, classifying, and documenting the requirements that impact the architecture. This agent is crucial to ensuring that the subsequent architecture design aligns with business goals.

To accomplish this task, the Analyst agent should possess capabilities to understand the SRS, which includes the following actions: 1) parse and structure the SRS to extract all requirements; 2) filter out architecture-significant requirements (ASRs); 3) classify functional and non-functional requirements, where functional requirements specify the system features and tasks, and non-functional requirements include quality attributes, resource constraints, and other relevant aspects. In addition to these actions, given that the SRS may be imperfect, the Analyst should also 4) identify potential risks in the SRS, such as ambiguous descriptions of system functionalities or conflicting quality attributes, and 5) communicate with stakeholders to address these issues and accordingly refine the SRS.

To support these actions, we propose that the Analyst agent should incorporate the following architectural knowledge: 1) methods for identifying and evaluating architecture-significant requirements, including their criticality and impact on design decisions; 2) understanding of the inherent trade-offs among various quality attributes, such as balancing performance, scalability, and security; 3) expertise in modeling non-functional requirements and mapping quality attributes to appropriate architectural tactics; 4) techniques for identifying and resolving risks in requirements, such as ambiguity, incompleteness, and conflicts; 5) knowledge of domain-specific constraints that influence architectural decisions.

\subsubsection{Modeler}

The Modeler agent is accountable for modeling the overall architecture of the system based on the refined requirements provided by the Analyst. It will generate 1) architectural decisions such as the selected technology stacks, architectural styles, and patterns; 2) conceptual logical views reflecting the systems' key domains; and 3) conceptual physical views outlining the system's overall deployment topology.

To fulfill these responsibilities, the Modeler should perform the following actions: 1) prioritize the non-functional requirements to facilitate trade-offs among quality attributes; 2) select the appropriate technology stack to meet the requirements; 3) choose suitable architecture styles and patterns to structure the system effectively; 4) identify domain components and their interrelationships to construct conceptual logical views; 5) allocate resources and plan the deployment model to construct conceptual physical views.

The Modeler agent also requires access to specific architectural knowledge to effectively perform its actions: 1) techniques for prioritizing non-functional requirements to balance trade-offs; 2) knowledge of technology stacks, including their capabilities in addressing quality requirements; 3) expertise in architectural styles and patterns, including their applicability and advantages; 4) methods for domain-driven design to identify domain components and establish relationships within the logical views; 5) strategies for resource allocation and deployment planning, including scalability considerations and mapping logical components to physical infrastructure.

\subsubsection{Designer}

The Designer agent is in charge of refining the conceptual views generated by the Modeler agent into detailed designs that serve as a foundation for subsequent code implementation. Key outputs of the Designer agent include: 1) class diagrams, defining the responsibilities, relationships, and interactions of system modules; 2) sequence diagrams, illustrating the communication and interaction flows between modules; and 3) deployment diagrams, specifying the resource allocation and deployment topology.

To fulfill its responsibilities, the Designer agent should execute the following actions: 1) define the responsibilities and boundaries of each module based on the established logical architecture; 2) design detailed interface specifications to facilitate module collaboration; 3) develop interaction models to outline module interaction mechanisms; 4) refine resource allocation plans to meet system constraints and performance goals.

Likewise, the Designer agent also requires specific architectural knowledge to effectively perform its actions: 1) techniques for modular design, adhering to design principles like SOLID principles~\cite{Martin2003}; 2) standards and best practices for interface design, ensuring compatibility and reusability; 3) methods for creating interaction models, such as sequence diagrams and collaboration diagrams, to represent communication flows; 4) strategies for resource optimization and allocation, particularly in distributed systems.

\subsubsection{Evaluator}

The Evaluator agent is tasked with rigorously assessing the architectural artifacts generated by other agents to ensure their alignment with the input SRS. Key outputs from the Evaluator agent include: 1) mismatch analysis reports, documenting discrepancies where the architectural outputs fail to meet the requirements specified in the SRS; 2) root cause analysis reports, identifying the underlying causes of the mismatches
and 3) refinement suggestions, providing actionable recommendations.

To fulfill these responsibilities, the Evaluator agent should perform the following actions: 1) evaluate the architectural outputs, including class, sequence, and deployment diagrams, to identify mismatches with the functional and non-functional requirements outlined in the SRS; 2) analyze the identified mismatches and trace back to determine their root causes, such as requirement misinterpretations, design errors, or resource misallocations; 3) formulate refinement suggestions to address the root causes and improve alignment with the system's goals and constraints.

The Evaluator agent also needs specific architectural knowledge to perform its actions: 1) techniques for validating architecture against functional and non-functional requirements, such as quality attribute scenarios and performance evaluations; 2) methods for root cause analysis, focusing on tracing mismatches back to requirement ambiguities or design flaws; 
3) different architectural evaluation standards, such as those specific to domains like real-time systems, cloud computing, and etc.

\section{Knowledge Extraction for MAAD}\label{Sec:Knowledge}

This section outlines the systematic extraction of architectural knowledge from three primary sources: existing projects, authoritative publications, and expert architects. We detail the knowledge extraction process for each source and subsequently demonstrate how the acquired knowledge empowers the Analyst, Modeler, Designer, and Evaluator agents to effectively fulfill their respective roles in the architectural design process.

\subsection{Knowledge from Existing Design}

To extract knowledge from existing projects, the process begins by gathering a comprehensive dataset of open-source or publicly available projects from various domains. Architecture recovery tools~\cite{garcia2011enhancing, zhang2023software} are employed to extract the architectural design of each selected project. Additionally, static or dynamic analysis tools~\cite{sonarqube} are utilized to evaluate the quality attributes of these projects, such as performance, scalability, and maintainability. By correlating the recovered architectures with the quality metrics, valuable knowledge is gained about the impact of architectural design. The knowledge can be applied to tasks like design trade-offs, resource allocation, and quality attribute optimization.

The extracted knowledge from existing projects can support the agents in several ways. For the Analyst agent, it could facilitate the identification and filtering of architecturally significant requirements by examining how similar requirements influenced decisions in other systems. The Modeler agent might leverage observations of architectural patterns and styles, along with their potential trade-offs in achieving desired quality attributes. The Designer agent can draw on extracted module designs and interface specifications derived from logical views, which may highlight design principles. The Evaluator agent can use documented mismatches between requirements and implementations, along with their associated resolutions, to inform their assessments of architectural alignment.

\subsection{Knowledge from Authoritative Literature}

Extracting knowledge from authoritative literature involves a comprehensive analysis of textbooks, academic papers, and industry standards that focus on architectural principles, patterns, and frameworks. NLP techniques can be applied to parse and extract structured information, such as definitions of architectural styles, trade-offs of quality attributes, and detailed examples of successful patterns. Furthermore, the literature provides frameworks for risk mitigation, modularization, and evaluation, which can be synthesized into actionable guidelines. Generalized methodologies for designing and evaluating architectures, such as ISO/IEC standards~\cite{iso2019software}, can also be extracted and adapted to support the agents.

The extracted knowledge could strengthen the capabilities of the agents in various ways. The Analyst agent benefits from frameworks for prioritizing non-functional requirements and techniques for identifying ASRs. The Modeler agent gains comprehensive descriptions of architectural styles and patterns, including their trade-offs and suitability for diverse contexts. The Designer agent utilizes guidelines for modular design, emphasizing design principles, as well as standards for interface design and communication modeling. The Evaluator agent relies on methodologies for root cause analysis and frameworks to evaluate architecture against quality attributes.

\subsection{Knowledge from Architecture Experts}

Knowledge extraction from experts involves engaging with seasoned practitioners through interviews, workshops, and surveys. Structured knowledge elicitation techniques, such as the Delphi method~\cite{linstone1975delphi}, can be employed to gather and organize insights. Experts will be queried about specific architectural challenges, emerging trends, and best practices based on their practical experience. Furthermore, recorded sessions will be transcribed and analyzed using NLP tools to identify recurring themes, actionable advice, and domain-specific recommendations. This process effectively captures the tacit knowledge and practical trade-offs that are often elusive in both literature and project data.

Expert insights significantly enhance the capabilities of the agents. Specifically, the Analyst agent relies on expert advice to elicit and resolve ambiguities in requirements and address domain-specific challenges. The Modeler agent benefits from tailored input on adapting architectural styles and patterns to unique scenarios. The Designer agent can gain practical feedback on emerging trends in module design, interface design, and resource optimization. The Evaluator agent can use expert strategies to identify subtle risks and refine evaluation processes. 

\section{Challenges and Opportunities}\label{Sec:Challenge}

Despite the rapid growth of AI4SE research and its extensive adoption, several challenges persist, presenting valuable avenues for future exploration. Based on our envisioned MAAD framework, we retrospectively and prospectively discuss those challenges and opportunities for future research.

\textbf{Explainability and Trustworthiness}: 
Despite the promising performance of LLMs on various software engineering and security tasks, such as code summarization~\cite{sun2025commenting, sun2024source, fang2024esale} and vulnerability detection~\cite{liu2024propertygpt, sun2024gptscan}, many previous studies~\cite{Shen2023chatgpt, chen2024security, ge2024demonstration, 2025-EliBadCode, 2025-KillBadCode} have raised concerns about trust in Artificial Intelligence Generated Content (AIGC).

A key strategy to improve reliability is effective knowledge injection into LLM-based multi-agent frameworks. Retrieval-Augmented Generation (RAG) improves decision-making by grounding agents in external knowledge, such as case studies and authoritative literature~\cite{gao2024retrieval}. Beyond retrieval, fine-tuning on collected architectural data can also enable agents to internalize best practices and industry standards~\cite{liu2022few}. Exploring how to effectively utilize or combine these strategies for architectural decision-making is a promising direction for future research.

\textbf{Coordination and Communication}: Each agent's role in the MAAD framework could encompass multiple LLMs-based agents to execute input tasks coordinately. The success of any multi-agent system hinges upon the seamless interplay of its constituent agents. This necessitates the design of robust and efficient mechanisms for information sharing, conflict resolution, and decision-making. 

\begin{itemize}[left=0pt]
    \item \textit{Information Sharing}: Agents must be able to effectively share relevant information, such as their current state, observations, goals, and plans, with other agents. This requires well-defined communication protocols and data formats to ensure that information is exchanged accurately and efficiently. Furthermore, mechanisms for filtering and prioritizing information are crucial to prevent information overload and ensure that agents focus on the most relevant data.

    \item \textit{Conflict Handling}: In many scenarios, multiple agents may have competing goals or may disagree on the best course of action. Robust conflict resolution mechanisms are therefore essential to ensure that the system can effectively navigate these challenges. This may involve negotiation protocols, arbitration mechanisms, or techniques for prioritizing the goals of different agents.

    \item \textit{Decision-Making}: Collective decision-making in a multi-agent system is inherently complex, necessitating a thoughtful evaluation of individual agents' preferences, priorities, and the specific features of the tasks at hand (e.g., prioritizing safety or stability). Effective decision-making mechanisms may include distributed consensus algorithms, voting protocols, or methods for aggregating individual preferences into a cohesive collective decision.
\end{itemize}

\textbf{Scalability and Dynamic}: 
In the context of complex architectural tasks, agents should exhibit dynamic workflows to effectively address evolving design requirements. While each agent follows a structured framework, the execution of their tasks can vary depending on the complexity and constraints of the problem. This flexibility allows agents to adapt their workflows dynamically, optimizing resource allocation and decision-making based on the specific demands of the task. Enabling agents to adjust their internal reasoning, decision strategies, or collaboration patterns based on contextual factors is a promising research direction. Additionally, future extensions of this framework could also explore dynamic agent configurations, where agent roles or specializations can evolve based on task complexity. Such a design may further enhance MAAD's adaptability in architectural design.

\section{Conclusion}\label{Sec:Conclusion}
In this paper, we envision the Multi-Agent Architecture Design (MAAD) framework, a novel approach to automating software architecture design by leveraging LLM-based agents. By simulating the roles of human architects, MAAD aims to enhance efficiency, scalability, and consistency in architectural tasks. The framework integrates knowledge from real-world systems, authoritative sources, and domain experts to guide the collaboration of four specialized agents, Analyst, Modeler, Designer, and Evaluator, toward producing high-quality architectures based on input Software Requirements Specifications.

The future roadmap for MAAD includes addressing critical challenges such as ensuring the explainability and trustworthiness of agent decisions, improving agent coordination through robust communication protocols, scaling the framework to dynamically adapt to complex design tasks, and effectively augmenting agents with extracted knowledge.

\bibliographystyle{ACM-Reference-Format}
\bibliography{FSE_ref}

\end{document}